\documentclass[twocolumn,showpacs,preprintnumbers,amsmath,amssymb]{revtex4}

\usepackage{graphicx}
\usepackage{dcolumn}
\usepackage{bm}

\newcommand{\as}{a\!\!\!/}
\newcommand{\As}{A\!\!\!/}
\newcommand{\ks}{k\!\!\!/}
\newcommand{\ps}{p\!\!\!/}

\begin{document}
\title{Electron's anomalous magnetic moment effects on electron-hydrogen elastic collisions in the presence of a circularly polarized laser field}
 \author{S. Elhandi} \author{S. Taj}\author{Y. Attaourti}\email{attaourti@ucam.ac.ma}
\affiliation{Laboratoire de Physique des Hautes
Energies et d'Astrophysique, Facult\'e des Sciences Semlalia, Universit\'e Cadi Ayyad Marrakech, BP : 2390, Maroc. }

\author{B. Manaut}
\affiliation{Universit\'e Sultan Moulay Slimane, Facult\'e Polydisciplinaire, Laboratoire Interdisciplinaire de Recherche en Sciences et  Techniques (LIRST)  BP : 523, 23000 B\'eni Mellal, Maroc. }

\author{L. Oufni}
\affiliation{Universit\'e Sultan Moulay Slimane, Facult\'e des Sciences et Techniques, D\'epartement de Physique, LPMM-ERM,  BP : 523, 23000 B\'eni Mellal, Maroc.}


\begin{abstract}
The effect of the electron's anomalous magnetic moment on the relativistic electronic dressing for the process of electron-hydrogen
atom elastic collisions is investigated. We consider a laser field with circular polarization and various electric field strengths. The Dirac-Volkov states taking into account this anomaly are used to describe the process in the first order of perturbation theory. The correlation between the terms coming from this anomaly and the electric field strength gives rise to new results, namely the strong dependence of the spinor part of the differential cross section (DCS) with respect to these terms. A detailed study has been devoted to the non relativistic regime as well as the moderate relativistic regime. Some aspects of this dependence as well as the dynamical behavior of the DCS in the relativistic regime have been addressed.

\end{abstract}

\pacs{34.50.RK, 34.80.Qb, 12.20.Ds}
\maketitle

\section{Introduction}
The value of the electron's magnetic moment is a fundamental
quantity in Physics. Its deviation from the value expected from
Dirac theory has given enormous impetus to the field of quantum
theory and especially to Quantum Electrodynamics (QED). It is
usually expressed in term of the $g$-factor, ( e.g for the electron $g=2$).
This result differs from the observed value by a small fraction
of a percent. The difference is the well known anomalous magnetic
moment, denoted $a$ and defined as : $a=(g-2)/2$. The
one-loop contribution to the anomalous magnetic moment of the
electron is found by calculating the vertex function. The
calculation is relatively straightforward \cite{1} and the
one-loop result is :
\begin{equation*}
a=\frac{\alpha}{2\pi}\simeq0.0011614
\end{equation*}
where $\alpha$ is the fine structure constant. This result was
first found by Schwinger \cite{2} in 1948. A recent experimental value for $a$
was obtained by Gabrielse \cite{3} :
\begin{equation*}
a=0.001159652180(73)
\end{equation*}
The state of the art status of QED predictions of the electron anomaly has been remarkably
reviewed by E. Remiddi \cite{4}. As for laser-assisted processes in Relativistic Atomic Physics,
the expectation of major advances in laser capabilities has
placed a new focus on the fundamentals of QED which
occupies a place of paramount importance among the theories used
in the formalism needed to obtain theoretical predictions for the intricate understanding of various fundamental processes.
QED has proven itself to be capable of remarkable
quantitative agreement between theoretical predictions and precise
laboratory measurements. When presently achievable intensities are around
$10^{22}W/cm^{2}$, electrons are so shaken that their velocity approaches the speed of light.
Therefore, the interactions between laser and matter become
relativistic.
Recently, relativistic laser-atom physics emerged as a new fertile reseach area. This is due to the newly opened possibility to submit atoms to ultra-intense pulses of infrared coherent radiation from lasers of various types. The dynamics of a free electron embedded within a constant amplitude classical field  has been addressed since the early years of quantum mechanics. In 1935, an exact expression for the wave function had been derived within the framework of the Dirac theory \cite{5}; see also \cite{6} for an overview of the case of an electron submitted to a short laser pulse. In the 1960s, the advent of laser devices has motivated theoretical studies related to QED in strong fields. These formal results were considered as being only of academic interest for many years. The state of affairs has significantly changed in the mid-1990s when it has been possible to make to collide a relativistic electron beam from a LINAC with a focused laser (Nd: Yag) radiation. Under such extreme conditions, it has been possible to evidence highly non-linear essentially relativistic QED processes such as a non linear Thomson and Compton scattering and also pair production [7,8,9,10,11]. A first focus issue in 1998 devoted to relativistic effects in strong fields appeared in Optics Express \cite{12} and in 2008, Strong Field Laser Physics \cite{13}  gave the main advances in this field as well as the references to the works of all major contributors.
A seminal thesis \cite{14}  for the first time addressed the study of laser-assisted second-order relativistic QED processes. This, we think will pave the way for a more accurate description of laser-assisted fundamental processes \cite{15}, \cite{16}.
 Our aim in this paper is to shed some light on a difficult and recently addressed description of laser-assisted processes that incorporate the electron anomaly. The process we study is the laser-assisted elastic collision of a Dirac-Volkov electron with a hydrogen atom. We focus on the relativistic electronic dressing with the addition of the electron anomaly. Some results are rather surprising bearing in mind the small value of $a$. In section 2, we present the formalism as well as the coefficients that intervene in the expression of the DCS. In section 3, we discuss the results we have obtained in the non relativistic, moderate relativistic and relativistic regimes. Atomic units are used throughout ($\hbar=e=m=1$) where $m$ denotes the electron mass and work with the metric tensor $g^{\mu\nu}=diag(1,-1,-1,-1)$.
\vspace{1in}
\\
\\
\begin{widetext}
\section{Theory}
The second-order Dirac equation for an electron with anomalous
magnetic moment (AMM) in the presence of an external
electromagnetic field is \cite{17} :

\begin{equation}
\left[(p-\frac{1}{c}A)^2-c^2-\frac{i}{2c}F_{\mu\nu}\sigma^{\mu\nu}+ia(\ps-\frac{\As}{c}+c)F_{\mu\nu}\sigma^{\mu\nu}\right]\psi(x)=0\label{1}
\end{equation}
where $\sigma^{\mu\nu}=\frac{1}{2}[\gamma^{\mu},\gamma^{\nu}]$,
$\gamma^{\mu}$ are the Dirac matrices and
$F_{\mu\nu}=\partial_{\mu}A_{\nu}-\partial_{\nu}A_{\mu}$ is the
electromagnetic field tensor. $A^{\mu}$ is the four-vector
potential, while $a=\kappa/4$, $\kappa$ is the electron's anomaly.
The Feynman slash notation is used throughout
$\ps=p_{0}\gamma^{0}-\mathbf{p}.\mathbf{\gamma}$ and
$\As=A_{0}\gamma^{0}-\mathbf{A}.\mathbf{\gamma}$. The term
$F_{\mu\nu}\sigma^{\mu\nu}$ stems from the fact that the electron
has a spin one half, the term multiplying $a$ is due to its
anomalous magnetic moment. It is possible to rewrite the exact
solution found by Y. I. Salamin \cite{18} as :
\begin{align}
\psi(x)=&\exp{\left\{-\left[\alpha\ks\As+\beta\ks+\delta\ps\ks\As\right]
\right\}}\frac{u(p,s)}{\sqrt{2VQ_0}}\exp{\left\{-i(p.x)+i\int_0^{kx}\left[\frac{A^2}{2c^2(k.p)}-\frac{(A.p)}{c(k.p)}\right]d\phi\right\}}\\\notag
       =&\exp{\left\{-\left[\alpha\ks\As+\beta\ks+\delta\ps\ks\As\right]\right\}}\frac{u(p,s)}{\sqrt{2VQ_0}}
         \exp{\left\{-i(q.x)-i\int_0^{kx}\frac{(A.p)}{c(k.p)}d\phi\right\}}
\end{align}
where
\begin{equation}
\alpha=- \frac{ac}{(k.p)}-\frac{1}{2c(k.p)}
\end{equation}
\begin{equation}
\beta=a\frac{A^2}{c(k.p)}
\end{equation}
\begin{equation}
\delta=\frac{a}{(k.p)}
\end{equation}
In the above equation the four-vector $q^{\mu}$ is given by :
\begin{equation}
q^{\mu}=p^{\mu}-\frac{A^{2}}{2c^{2}(k.p)}k^{\mu}
\end{equation}
The four-vector $k^{\mu}=(\frac{\omega}{c},\mathbf{k})$ is the
four-vector of the circularly polarized laser field
$A^{\mu}=a_{1}^{\mu}\cos\phi+a_{2}^{\mu}\sin\phi$, $\phi=k.x$.
Note that $k^{\mu}$ is such that $k^{2}=0$ and $k_{\mu}A^{\mu}=0$
implying that we are working in the Lorentz gauge. One has
\begin{equation}
q^{\mu}q_{\mu}=\frac{Q^{2}}{c^{2}}-\mathbf{q}^{2}=\left(1-\frac{\overline{A^{2}}}{c^{4}}\right)c^{2}={m^{*}}^{2}c^{2}
\end{equation} where $m^{*}$ is the effective mass that the
electron acquires when embedded within a laser field :
\begin{equation}
m^{*}=\left(1-\frac{\overline{A^{2}}}{c^{4}}\right)^{\frac{1}{2}}
\end{equation}
We are considering laser intensities such that the resulting
ponderomotive force is comparable to its rest mass, it is then
possible to retain only terms of order one in the expansion of the
first exponential in Eq.2 and find :
\begin{equation}
\psi(x)=\left[1-[\alpha\ks\As+\beta\ks+\delta\ps\ks\As\right]\frac{u(p,s)}{\sqrt{2VQ_0}}
         \exp{\left\{-i(q.x)-i\int_0^{kx}\frac{(A.p)}{c(k.p)}d\phi\right\}}
\end{equation}
This expression differs formally from that found by various
authors \cite{19} but is actually equivalent. The transition matrix
element corresponding to the process of laser assisted
electron-atomic hydrogen is given by :
\begin{equation}
S_{fi}=-i\int
dt\left<\psi_{p_{f}}(x_{1})\phi_{f}(x_{2})|V_{d}|\psi_{p_{i}}(x_{1})\phi_{i}(x_{2})\right>
\end{equation}
The explicit expression of the wave function of the hydrogen atom
$\phi(x_{2})$ for the fundamental state (spin up) can be found in
\cite{20} and reads :
\begin{eqnarray}
\phi(x_{2})=\frac{1}{\sqrt{4\pi}} \left\{\begin{array}{c}
ig(x_{2})\\
0\\
f(x_{2})\cos\theta\\
f(x_{2})\cos\theta e^{i\phi}\end{array}\right\}
\end{eqnarray}
where $g(x_{2})$ is given by :
\begin{equation}
g(x_{2})=(2Z)^{\gamma+\frac{1}{2}}\sqrt{\frac{1+\gamma}{2\Gamma(1+2\gamma)}}e^{-Zx_{2}}x_{2}^{\gamma-1}=N_{g}e^{-Zx_{2}}x_{2}^{\gamma-1}
\end{equation}
and
\begin{equation}
f(x_{2})=-(2Z)^{\gamma+\frac{1}{2}}\sqrt{\frac{1+\gamma}{2\Gamma(1+2\gamma)}}e^{-Zx_{2}}x_{2}^{\gamma-1}\frac{(1-\gamma)}{Z\alpha}=N_{f}e^{-Zx_{2}}x_{2}^{\gamma-1}
\end{equation}
The instantaneous interaction potential $V_{d}$ is given by :
\begin{equation}
V_{d}=\frac{1}{x_{12}}-\frac{Z}{x_{1}}
\end{equation}
where $\mathbf{x}_{12}=|\mathbf{x}_{1}-\mathbf{x}_{2}|$.
$\mathbf{x}_{1}$ is the electron coordinates and $\mathbf{x}_{2}$
are the atomic electron coordinates. If we replace all wave
functions in Eq.10, the transition matrix element $S_{fi}$
becomes :
\begin{equation}
S_{fi}=-i\sum_{n=-\infty}^{+\infty}\frac{2\pi\delta(Q_{f}-Q_{i}-n\omega)}{2V\sqrt{Q_{i}Q_{f}}}H(\mathbf{\Delta}_{n})\left|\overline{u}(p_{f},s_{f})\Gamma_{n}u(p_{i},s_{i})\right|
\end{equation}
where the expression of $H(\mathbf{\Delta}_{n})$ has already been
derived in a previous work \cite{21},where
$\mathbf{\Delta}_{n}=\left|\mathbf{q}_{f}-\mathbf{q}_{i}-n\mathbf{k}\right|$
is the momentum transfer with the net exchange of $n$ photons.We
have obtained for $H(\mathbf{\Delta}_{n})$ the following
analytical expression :
\begin{equation}
H(\mathbf{\Delta}_{n})=-4\pi(N_{g}^{2}+N_{f}^{2})\Gamma(2\gamma+1)\left(\frac{1}{(2Z)^{2\gamma+1}\Delta_{n}^{2}}-\frac{\sin(2\gamma\phi)}{2\gamma\lambda^{2\gamma}\Delta_{n}^{3}}\right)
\end{equation}
with
\begin{equation}
\lambda=\sqrt{(2Z)^{2}+\Delta_{n}^{2}} \quad \text{and}\quad
\phi=\arctan\left(\frac{\Delta_{n}}{2Z}\right)
\end{equation}
However, the novelty in the various stages of the calculations is
contained in the term
$\left|\overline{u}(p_{f},s_{f})\Gamma_{n}\overline{u}(p_{i},s_{i})\right|$,
where
\begin{align}
\Gamma_{n}=&\overline{R}(p_{f})\gamma^{0}R(p_{i})\\\notag
          =&\left[1-(\alpha_{f}\As\ks+\beta_{f}\ks+\delta_{f}\As\ks\ps_{f})\right]\gamma^{0}\left[1-(\alpha_{i}\ks\As+\beta_{i}\ks+\delta_{i}\ps_{i}\ks\As)\right]\\\notag
          =&C_{0}+C_{1}\cos\phi+C_{2}\sin\phi+C_{3}\cos2\phi+C_{4}\sin2\phi
\end{align}
These coefficients can be obtained using Reduce \cite{22}. We give
their analytical expressions :
\begin{align}
C_{0}=&(2*(2*\as_{1}\ks\ps_{i}*a_{1}.p_{f}*\delta_{f}*\delta_{i}*\omega-2*\as_{1}\ks\ps_{f}*a_{1}.p_{i}*\delta_{f}*\delta_{i}*\omega+2*\as_{1}\ks\gamma_{0}*a_{1}.p_{i}*k.p_{f}\\\notag
     &*c*\delta_{f}*\delta_{i}-2*\as_{1}\ks\gamma_{0}*a_{1}.p_{f}*k.p_{i}*c*\delta_{f}*\delta_{i}-2*\as_{1}\ks*a_{1}.p_{i}*\alpha_{f}*\delta_{i}*\omega+2*\as_{1}\ks\\\notag
     &*a_{1}.p_{f}*\alpha_{i}*\delta_{f}*\omega+2*\as_{2}\ks\ps_{i}*a_{2}.p_{f}*\delta_{f}*\delta_{i}*\omega-2*\as_{2}\ks\ps_{f}*a_{2}.p_{i}*\delta_{f}*\delta_{i}*\omega+2\\\notag
     &*\as_{2}\ks\gamma_{0}*a_{2}.p_{i}*k.p_{f}*c*\delta_{f}*\delta_{i}-2*\as_{2}\ks\gamma_{0}*a_{2}.p_{f}*k.p_{i}*c*\delta_{f}*\delta_{i}-2*\as_{2}\ks*a_{2}.p_{i}*\alpha_{f}\\\notag
     &*\delta_{i}*\omega+2*\as_{2}\ks*a_{2}.p_{f}*\alpha_{i}*\delta_{f}*\omega-2*\ks\ps_{i}*\alpha_{f}*A^{2}*\delta_{i}*\omega+2*\ks\ps_{f}\ps_{i}*A^{2}*\delta_{f}*\delta_{i}\\\notag
     &*\omega-2*\ks\ps_{f}\gamma_{0}*k.p_{i}*A^{2}*c*\delta_{f}*\delta_{i}+2*\ks\ps_{f}*\alpha_{i}*A^{2}*\delta_{f}*\omega-2*\ks\gamma_{0}\ps_{i}*k.p_{f}*A^{2}\\\notag
     &*c*\delta_{f}*\delta_{i}+2*\ks\gamma_{0}*k.p_{i}*\alpha_{f}*A^{2}*c*\delta_{i}-2*\ks\gamma_{0}*k.p_{f}*\alpha_{i}*A^{2}*c*\delta_{f}-\ks\gamma_{0}*\beta_{f}\\\notag
     &*c-2*\ks*\alpha_{f}*\alpha_{i}*A^{2}*\omega+2*\ks*\beta_{f}*\beta_{i}*\omega-\gamma_{0}\ks*\beta_{i}*c+\gamma_{0}*c))/(2*c)\\
C_{1}=&(2*(-\as_{1}\ks\ps_{f}\gamma_{0}*c*\delta_{f}+2*\as_{1}\ks\ps_{f}*\beta_{i}*\delta_{f}*\omega-2*\as_{1}\ks\gamma_{0}*k.p_{f}*\beta_{i}*c*\delta_{f}-\as_{1}\ks\gamma_{0}\\\notag
      &*\alpha_{f}*c+2*\as_{1}\ks*\alpha_{f}*\beta_{i}*\omega+2*\ks\as_{1}*\alpha_{i}*\beta_{f}*\omega-2*\ks\ps_{i}\as_{1}*\beta_{f}*\delta_{i}*\omega+2*\ks\gamma_{0}\as_{1}\\\notag
      &*k.p_{i}*\beta_{f}*c*\delta_{i}-\gamma_{0}\ks\as_{1}*\alpha_{i}*c-\gamma_{0}\ps_{i}\ks\as_{1}*c*\delta_{i}))/(2*c)\\
C_{2}=&(2*(-\as_{2}\ks\ps_{f}\gamma_{0}*c*\delta_{f}+2*\as_{2}\ks\ps_{f}*\beta_{i}*\delta_{f}*\omega-2*\as_{2}\ks\gamma_{0}*k.p_{f}*\beta_{i}*c*\delta_{f}\\\notag
      &-\as_{2}\ks\gamma_{0}*\alpha_{f}*c+2*\as_{2}\ks*\alpha_{f}*\beta_{i}*\omega+2*\ks\as_{2}*\alpha_{i}*\beta_{f}*\omega-2*\ks\ps_{i}\as_{2}\\\notag
      &*\beta_{f}*\delta_{i}*\omega+2*\ks\gamma_{0}\as_{2}*k.p_{i}*\beta_{f}*c*\delta_{i}-\gamma_{0}\ks\as_{2}*\alpha_{i}*c-\gamma_{0}\ps_{i}\ks\as_{2}*c*\delta_{i})/(2*c)\\
C_{3}=&(4*\as_{1}\ks\ps_{i}*a_{1}.p_{f}*\delta_{f}*\delta_{i}*\omega-\as_{1}\ks\ps_{f}*a_{1}.p_{i}*\delta_{f}*\delta_{i}*\omega+\as_{1}\ks\gamma_{0}*a_{1}.p_{i}*c\\\notag
      &*k.p_{f}*\delta_{f}*\delta_{i}-\as_{1}\ks\gamma_{0}*a_{1}.p_{f}*k.p_{i}*c*\delta_{f}*\delta_{i}-\as_{1}\ks*a_{1}.p_{i}*\alpha_{f}*\delta_{i}*\omega+\as_{1}\ks\\\notag
      &*a_{1}.p_{f}*\alpha_{i}*\delta_{f}*\omega-\as_{2}\ks\ps_{i}*a_{2}.p_{f}*\delta_{f}*\delta_{i}*\omega+\as_{2}\ks\ps_{f}*a_{2}.p_{i}*\delta_{f}*\delta_{i}*\omega-\as_{2}\ks\gamma_{0}\\\notag
      &*a_{2}.p_{i}*k.p_{f}*c*\delta_{f}*\delta_{i}+\as_{2}\ks\gamma_{0}*a_{2}.p_{f}*k.p_{i}*c*\delta_{f}*\delta_{i}+\as_{2}\ks*a_{2}.p_{i}*\alpha_{f}*\delta_{i}*\omega\\\notag
      &-\as_{2}\ks*a_{2}.p_{f}*\alpha_{i}*\delta_{f}*\omega))/(2*c)\\
C_{4}=&(4*(\as_{1}\ks\ps_{i}*a_{2}.p_{f}*\delta_{f}*\delta_{i}*\omega-\as_{1}\ks\ps_{f}*a_{2}.p_{i}*\delta_{f}*\delta_{i}*\omega+\as_{1}\ks\gamma_{0}*a_{2}.p_{i}*k.p_{f}*c\\\notag
      &*\delta_{f}*\delta_{i}-\as_{1}\ks\gamma_{0}*a_{2}.p_{f}*k.p_{i}*c*\delta_{f}*\delta_{i}-\as_{1}\ks*a_{2}.p_{i}*\alpha_{f}*\delta_{i}*\omega+\as_{1}\ks*a_{2}.p_{f}*\alpha_{i}\\\notag
      &*\delta_{f}*\omega+\as_{2}\ks\ps_{i}*a_{1}.p_{f}*\delta_{f}*\delta_{i}*\omega-\as_{2}\ks\ps_{f}*a_{1}.p_{i}*\delta_{f}*\delta_{i}*\omega+\as_{2}\ks\gamma_{0}*a_{1}.p_{i}*c\\\notag
      &*k.p_{f}*\delta_{f}*\delta_{i}-\as_{2}\ks\gamma_{0}*a_{1}.p_{f}*k.p_{i}*c*\delta_{f}*\delta_{i}-\as_{2}\ks*a_{1}.p_{i}*\alpha_{f}*\delta_{i}*\omega+\as_{2}\ks\\\notag
      &*a_{1}.p_{f}*\alpha_{i}*\delta_{f}*\omega))/(2*c)
\end{align}

The coefficients $C_{i}$'s contain all the information
about the effect of the electron's AMM since they depend on
$\kappa=4a$. Therefore, it is to be expected that this
afore-mentioned effect will be of a crucial importance since both
$\kappa$ and electric field strength are correlated in the
expression of the five coefficients $C_{i}$. We now introduce the well known relations involving ordinary Bessel functions :
\begin{eqnarray}
\left\{\begin{array}{c}
1\\
\cos(\phi)\\
\sin(\phi)\\
\cos(2\phi)\\
\sin(2\phi)\end{array}\right\}e^{-iz\sin(\phi-\phi_0)}=\sum_{n=-\infty}^{\infty}\left\{\begin{array}{c}
B_{0n}\\
B_{1n}\\
B_{2n}\\
B_{3n}\\
B_{4n}\end{array}\right\}e^{-in\phi},\label{23}
\end{eqnarray}
with
\begin{eqnarray}
\left\{\begin{array}{c}
B_{0n}\\
B_{1n}\\
B_{2n}\\
B_{3n}\\
B_{4n}\end{array}\right\}=\left\{\begin{array}{c}
J_n(z)e^{in\phi_0}\\
(J_{n+1}(z)e^{i(n+1)\phi_0}+J_{n-1}(z)e^{i(n-1)\phi_0})/2\\
(J_{n+1}(z)e^{i(n+1)\phi_0}-J_{n-1}(z)e^{i(n-1)\phi_0})/2i\\
(J_{n+2}(z)e^{i(n+2)\phi_0}+J_{n-2}(z)e^{i(n-2)\phi_0})/2\\
(J_{n+2}(z)e^{i(n+2)\phi_0}-J_{n-2}(z)e^{i(n-2)\phi_0})/2i\end{array}\right\}.
\label{24}
\end{eqnarray}
The product of terms given by Eq.18 can be written as :
\begin{equation}
\Gamma_{n}=\sum^{+\infty}_{n=-\infty}\left[C_{0}B_{0n}+C_{1}B_{1n}+C_{2}B_{2n}+C_{3}B_{3n}+C_{4}B_{4n}\right]
\end{equation}
Proceeding along the lines of standard QED calculations \cite{8}
we obtain for the formal DCS expression in the presence of
circularly polarized laser field and taking into account the
anomalous
magnetic moment of the electron :\\
\begin{equation}
\left.\frac{d\sigma}{d\Omega_{f}}\right|_{Q_{f}=Q_{i}+n\omega}=\left.\sum_{n=\infty}^{\infty}\frac{1}{(4\pi
c^{2})^{2}}\frac{|\mathbf{q}_{f}|}{|\mathbf{q}_{i}|}\frac{1}{2}\sum_{s_{i}}\sum_{s_{f}}\left|M_{fi}^{(n)}\right|^{2}\left|H(\mathbf{\Delta}_{n})\right|^{2}\right|_{Q_{f}=Q_{i}+n\omega}
\end{equation}
where
\begin{equation}
\frac{1}{2}\sum_{s_{i}}\sum_{s_{f}}\left|M_{fi}^{(n)}\right|^{2}=\frac{1}{2}Tr\left\{(\ps_{f}c+c^{2})\Gamma_{n}(\ps_{i}c+c^{2})\overline{\Gamma}_{n}\right\}
\end{equation}
and
\begin{equation}
\overline{\Gamma}_{n}=\gamma^{0}\Gamma_{n}^{\dagger}\gamma^{0}
\end{equation}
\end{widetext}
\section{results and discussions}
The geometry chosen is $\theta_{i}=\phi_{i}=45^{\circ}$ for the
incident electron while for the scattered electron
$\phi_{f}=90^{\circ}$ and the angle $\theta_{f}$ varies from
$-180^{\circ}$ to $180^{\circ}$. For all the angular
distributions of the DCSs, we maintain the same choice of the
laser angular frequency, that is, $\omega=0.0043$ $a.u$ which
corresponds to a near-infrared Neodymium laser. Also, for the
investigation of the behaviour of the various DCSs with respect
to the electric field strength $\varepsilon$ and the relativistic
parameter $\gamma$,
$\omega$ is fixed at the same value.\\
The dependence of the DCSs with respect to the laser frequency is
investigated beginning by a frequency $\omega_{min}=0.04$ $a.u$
to $\omega_{max}=0.1$ $a.u$.\\
We define some abbreviations that will be useful and simplify the
readability of our text,
$\left(d\sigma/d\Omega_f\right)^{RNL}$ means relativistic
DCS without a laser field while
$\left(d\sigma/d\Omega_f\right)^{RL}$ means relativistic DCS
with a laser field, $\left(d\sigma/d\Omega_f\right)^{NRL}$
is the non relativistic DCS with a laser field and
$\left(d\sigma/d\Omega_f\right)^{RL}_{AMM}$ is DCS with
anomalous magnetic moment in the presence of a laser field.

\subsection{The non relativistic regime.}
For an electron with a low kinetic energy and a moderate field
strength, typically $E_{c}=100$ $a.u$ or ($E_{c}=2700$ $ev$) and
a field strength $\varepsilon=0.05$ $a.u$, the non dressed
momentum coordinates $(\mathbf{p}_{i}, \mathbf{p}_{f})$ and
$(\mathbf{q}_{i}, \mathbf{q}_{f})$ are relatively close.\\
We have carried out numerical simulations for
$\varepsilon=0.05$ $a.u$ and $\gamma=1.0053$. Since the DCS s are
 sensitive to the number $n$ of photons exchanged, we have
used $\left(d\sigma/d\Omega_f\right)^{RNL}$ as a reference to see
how the others evolve with increasing values of $n$. For
$n=\pm100$ photons, the figure obtained is not physically sound
since the non relativistic, the
$\left(d\sigma/d\Omega_f\right)^{RL}$ and
$\left(d\sigma/d\Omega_f\right)^{RL}_{AMM}$ are close but a small
difference appears at the peaks located near
$\theta_{f}=33^{\circ}$. In other words, $n$ must be increased.
For $n=\pm400$ photons, a newly found result (as far as we know)
is the violation of the pseudo sum rule \cite{23}: the summed DCS
must
always converge toward the $\left(d\sigma/d\Omega_f\right)^{RNL}$ and therefore must be less than the latter.\\
\begin{figure}
\includegraphics[angle=0,width=2.5 in,height=3 in]{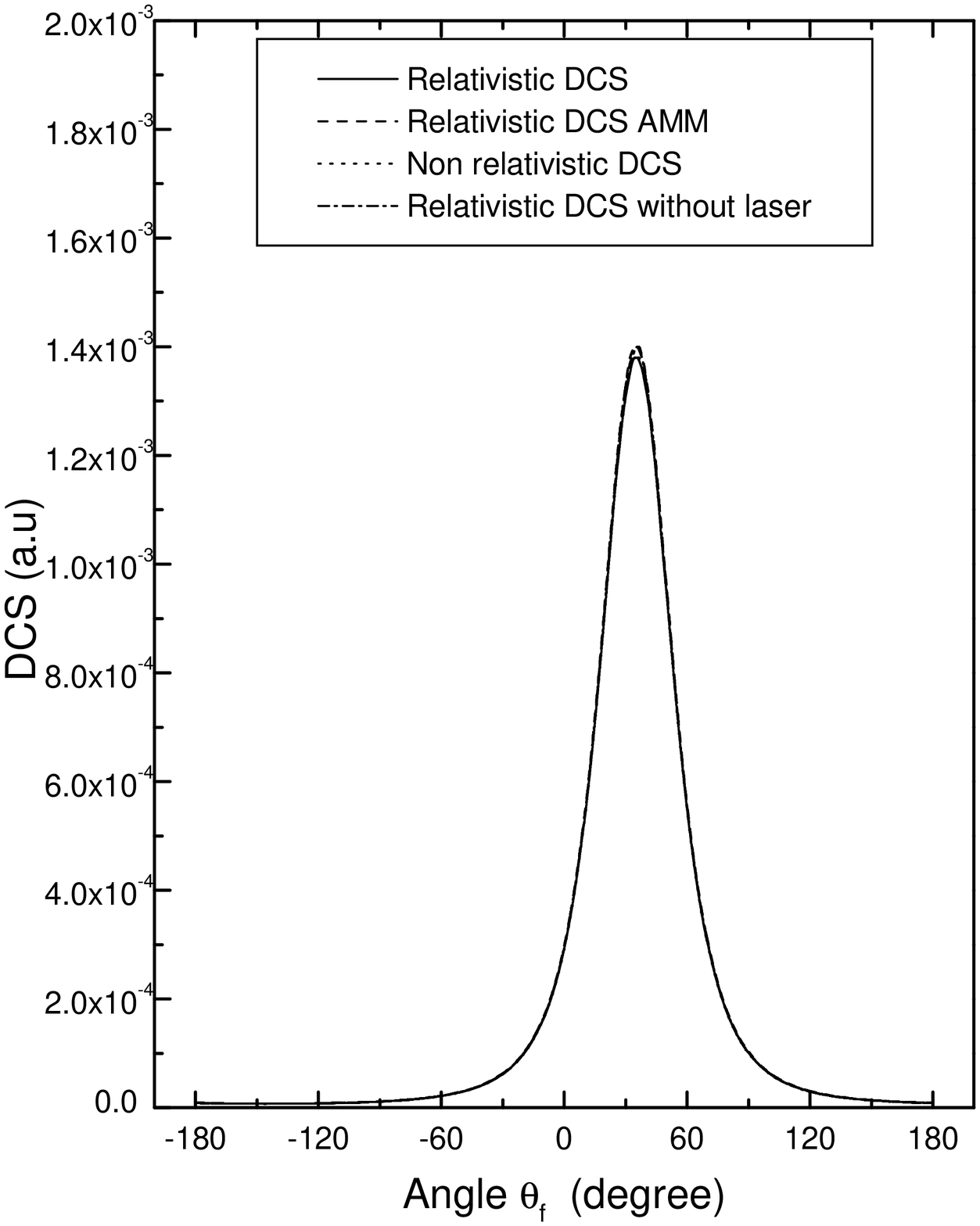}
\caption{ The various DCSs as a function of the
angle $\theta_{f}$ in degrees for an electrical field strength of
$\varepsilon=0.05$ $a.u$ and a relativistic parameter
$\gamma=1.0053$. The corresponding number of photons exchanged is
$\pm1200$.}
\end{figure}
The effect of the AMM of the electron plays a key role
for this behaviour. But this violation has to be ascertained by
increasing the number $n$ of photons. For $n=\pm1200$ photons,
upwards, this violation is confirmed meaning that even at low
kinetic incident energies and moderate field strength, the effect
of the AMM of the electron begins to be distinguished even if
this effect is small. For this number $n$ of photons the
$\left(d\sigma/d\Omega_f\right)^{RL}_{AMM}$ is higher at
the peak than the $\left(d\sigma/d\Omega_f\right)^{RNL}$
while $\left(d\sigma/d\Omega_f\right)^{NRL}$ and
$\left(d\sigma/d\Omega_f\right)^{RL}$ are close to each
other which was to be expected since in the non relativistic
regime, spin effects are
small. This is shown in Fig.1.\\
If we maintain the value of the relativistic parameter
$\gamma=1.0053$ and we increase the value of the electric field
strength $\varepsilon$ from $0.05$ $a.u$ to $0.1$ $a.u$, the
violation of the pseudo sum-rule is still present while the
difference between the
$\left(d\sigma/d\Omega_f\right)^{RL}_{AMM}$ and the
$\left(d\sigma/d\Omega_f\right)^{RL}$ is more pronounced
compared to the previous case where $\varepsilon=0.05$ $a.u$.
This is shown in Fig.2 where $n$ is also taken to have the value
$\pm1200$. At the peaks of the DCSs the difference between these
two is roughly equal to $6$ percent.\\
\begin{figure}[h]
\includegraphics[angle=0,width=2.5 in,height=3 in]{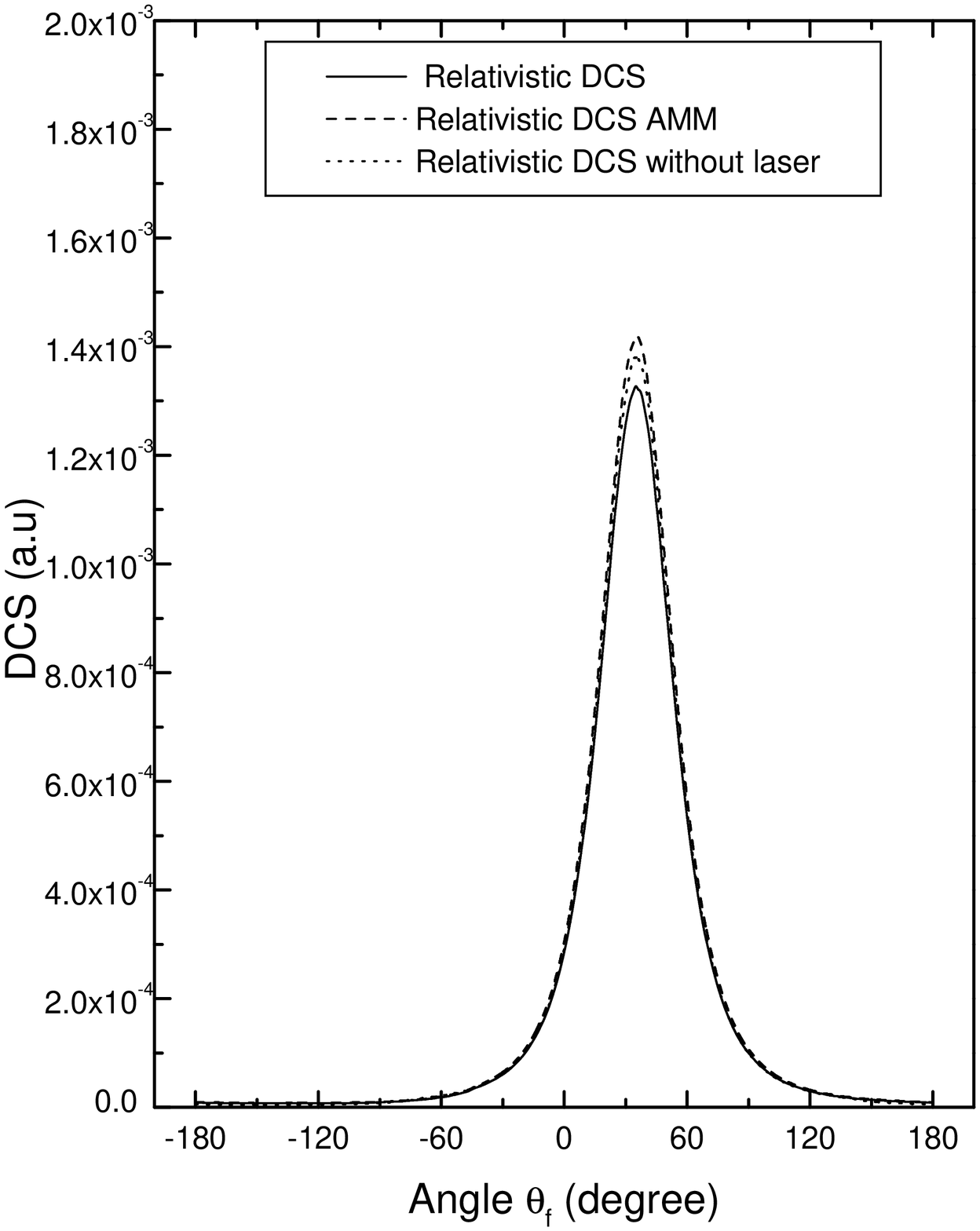}
\caption{ The various relativistic DCSs as a function of the angle
$\theta_{f}$ in degrees for an electrical field strength of
$\varepsilon=0.1$ $a.u$ and a relativistic parameter
$\gamma=1.0053$. The corresponding number of photons exchanged is
$\pm1200$.}
\end{figure}
\begin{figure}[h]
\includegraphics[angle=0,width=2.5 in,height=3 in]{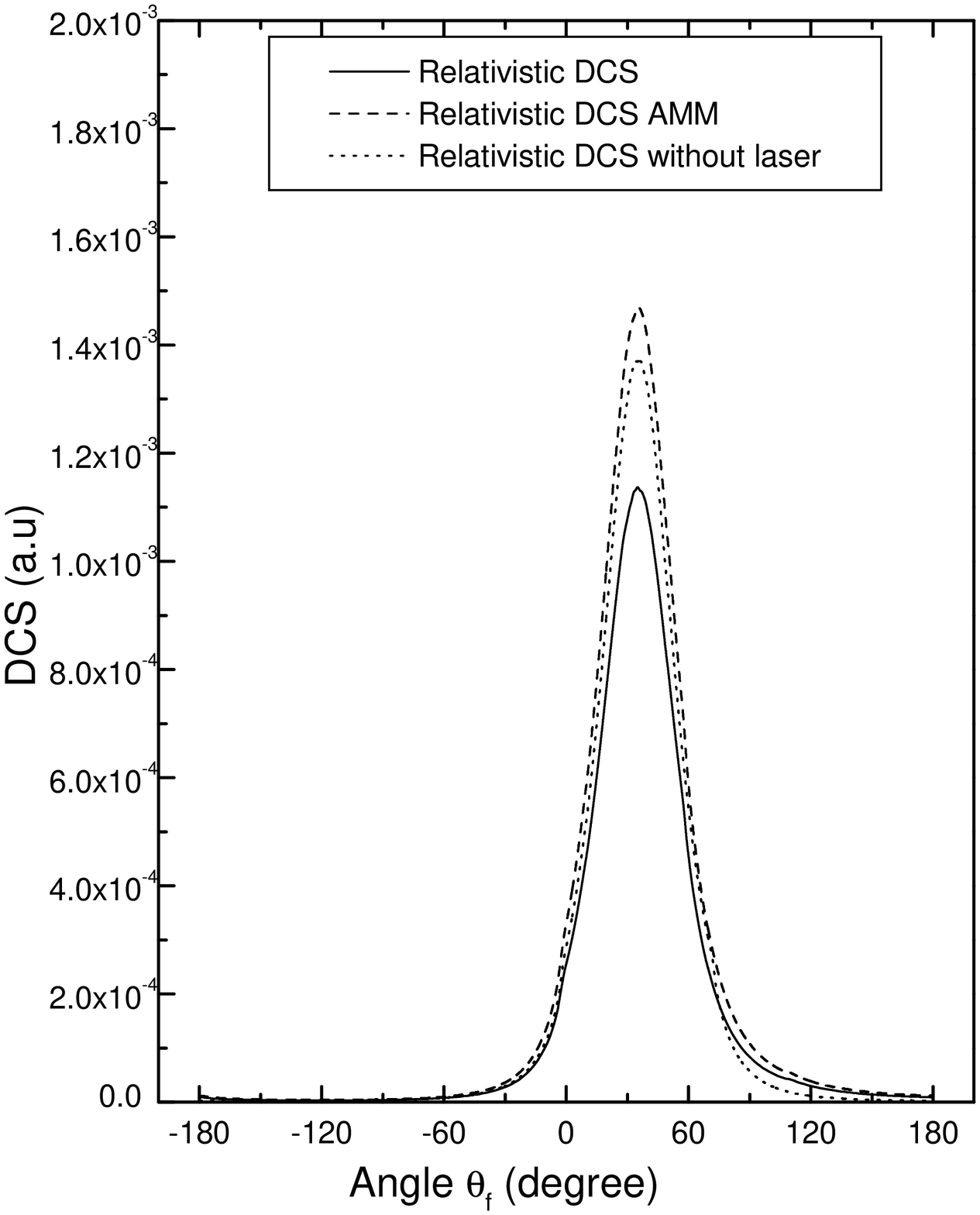}
\caption{ The various relativistic DCSs as a function of the angle
$\theta_{f}$ in degrees for an electrical field strength of
$\varepsilon=0.2$ $a.u$ and a relativistic parameter
$\gamma=1.0053$. The corresponding number of photons exchanged is
$\pm1200$.}
\end{figure}
As the electric field strength is increased from $\varepsilon=0.1$
$a.u$ to $\varepsilon=0.2$ $a.u$, the physical insights mentioned
are the same with a more noticeable difference between
$\left(d\sigma/d\Omega_f\right)^{RL}_{AMM}$ and
$\left(d\sigma/d\Omega_f\right)^{RL}$ of $36$
percent also at the peaks. This is shown in Fig.3.\\
\begin{figure}[h]
\includegraphics[angle=0,width=2.5 in,height=3 in]{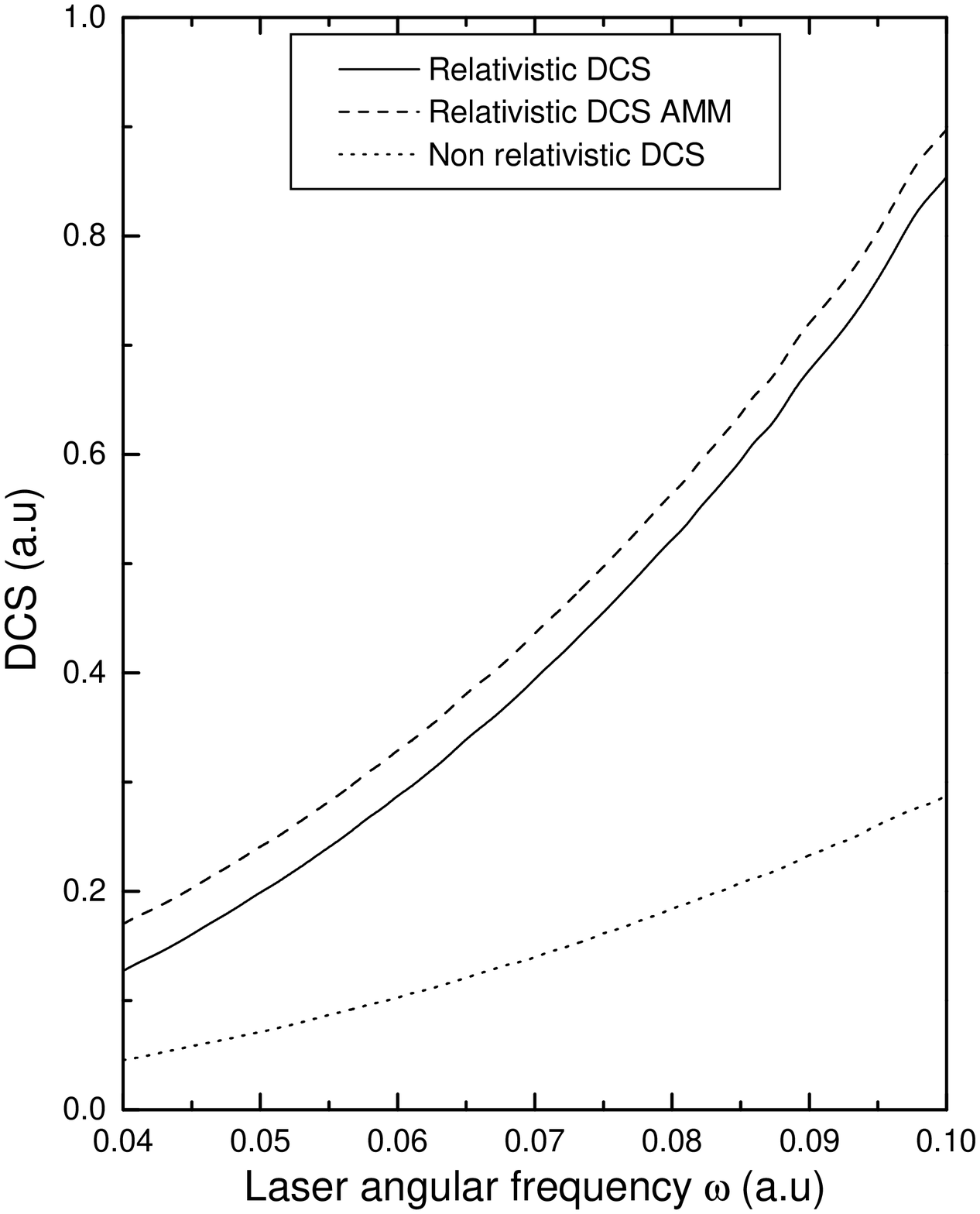}
\caption{ The various relativistic DCSs scaled in $10^{-5}$ as a function of $\omega$
for an electrical field strength of $\varepsilon=0.2$ $a.u$, a
relativistic parameter $\gamma=1.0053$ and an angle
$\theta_{f}=135^{\circ}$. The corresponding number of photons
exchanged is $\pm100$.}
\end{figure}
An interesting behaviour (with respect to the laser angular
frequency $\omega$ varying from $0.04$ $a.u$ to $0.1$ $a.u$ and
for $n=\pm100$ photons in the non relativistic regime
($\gamma=1.0053$) and the same angular momentum coordinates)
emerges with increasing $\varepsilon$, particularly for
$\varepsilon=0.2$ $a.u$ where
$\left(d\sigma/d\Omega_f\right)^{RL}_{AMM}$ is similar in
shape as $\left(d\sigma/d\Omega_f\right)^{RL}$ but always
higher. This advocates the fact that the term
$1/2\sum_{s_{i}}\sum_{s_{f}}\left|M_{fi}^{(n)}\right|^{2}$
is very sensitive to the variation of $\varepsilon$ and this fact
has to remain true for the relativistic regime. This is shown in
Fig.4.\\
It is then necessary to ascertain this dependence with respect to
the electric field strength by varying it from $\varepsilon=0.05$
$a.u$ to $\varepsilon=1$ $a.u$ while retaining all these same
other parameters. From $\varepsilon=0.05$ $a.u$ to
$\varepsilon=0.2$ $a.u$ the differences between the two
interesting DCS are visible and begin to separate drastically up
to $\varepsilon=1$ $a.u$ where,
$\left(d\sigma/d\Omega_f\right)^{RL}_{AMM}\approx11\left(d\sigma/d\Omega_f\right)^{RL}$.
\begin{figure}[h]
\includegraphics[angle=0,width=2.5 in,height=3 in]{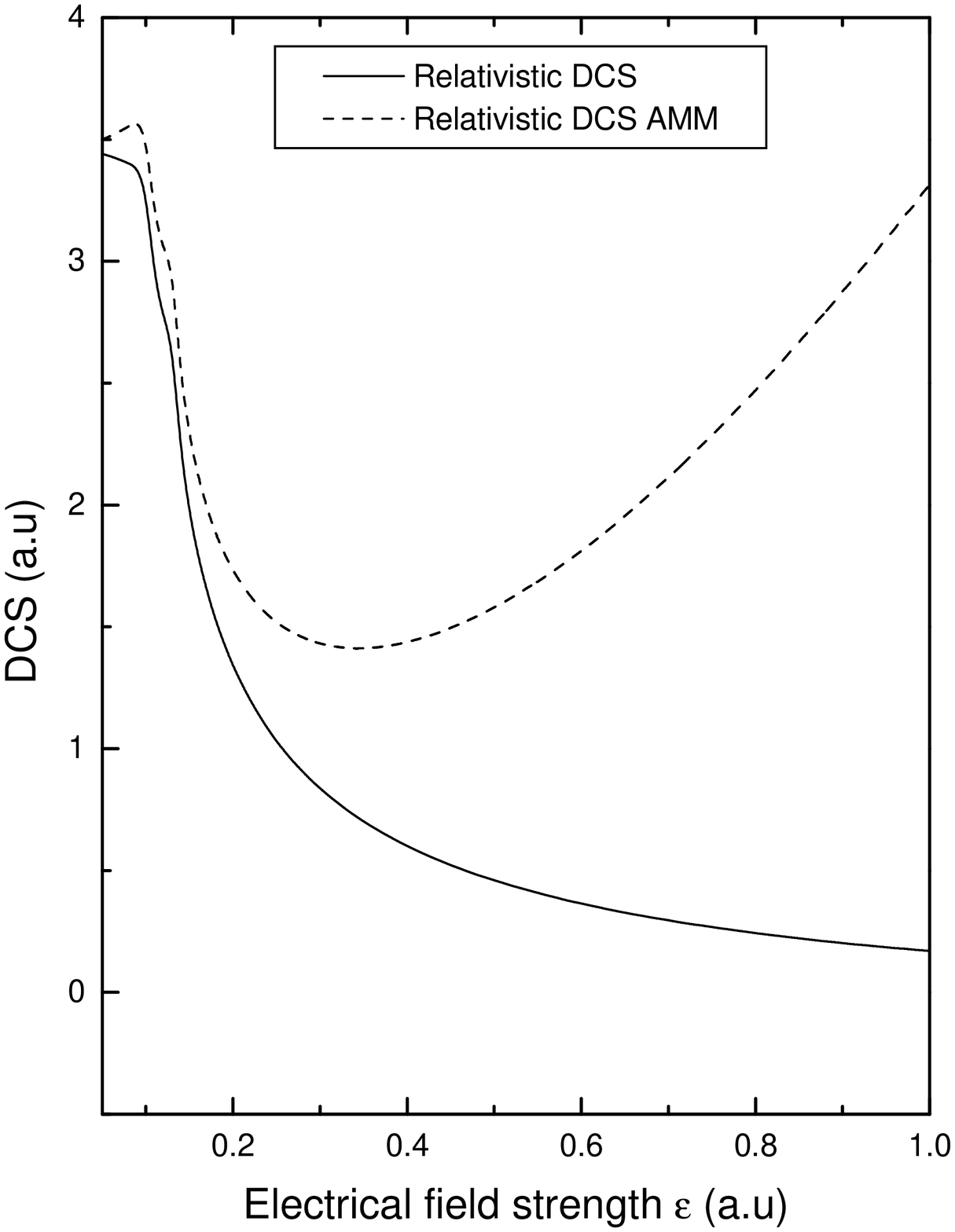}
\caption{ The two interesting DCSs scaled in $10^{-5}$ as a function of electrical
field strength of $\varepsilon$ for a relativistic parameter
$\gamma=1.0053$ and an angle $\theta_{f}=135^{\circ}$. The
corresponding number of photons exchanged is $\pm1000$.}
\end{figure}
Fig.5 clearly shows this behaviour as well as the strong
dependence with respect to the electric field strength of
$\left(d\sigma/d\Omega_f\right)^{RL}_{AMM}$.\\
Having given sound evidence of the role of the electric field
strength, we now turn to the dynamical behaviour of the various
DCSs with respect to the relativistic parameter $\gamma$ that is
taken to vary from $\gamma=1.0053$ to $\gamma=1.025$ i.e
maintaining the various simulations within the framework of the
non relativistic regime. For $\varepsilon=0.05$ $a.u$ and
$n=\pm200$ the difference between the two relevant DCSs is small.
This is also the case for $n=\pm400$. Moreover, increasing
$\varepsilon$ from $0.05$ $a.u$ to $0.1$ $a.u$ and fixing $n$ to
vary between $\pm1000$ gives a noticeable difference between the
two commonly studied DCSs while at the same time confirming the
violation of
the pseudo sum-rule. This is shown in Fig.6.\\
\begin{figure}[h]
\includegraphics[angle=0,width=2.5 in,height=3 in]{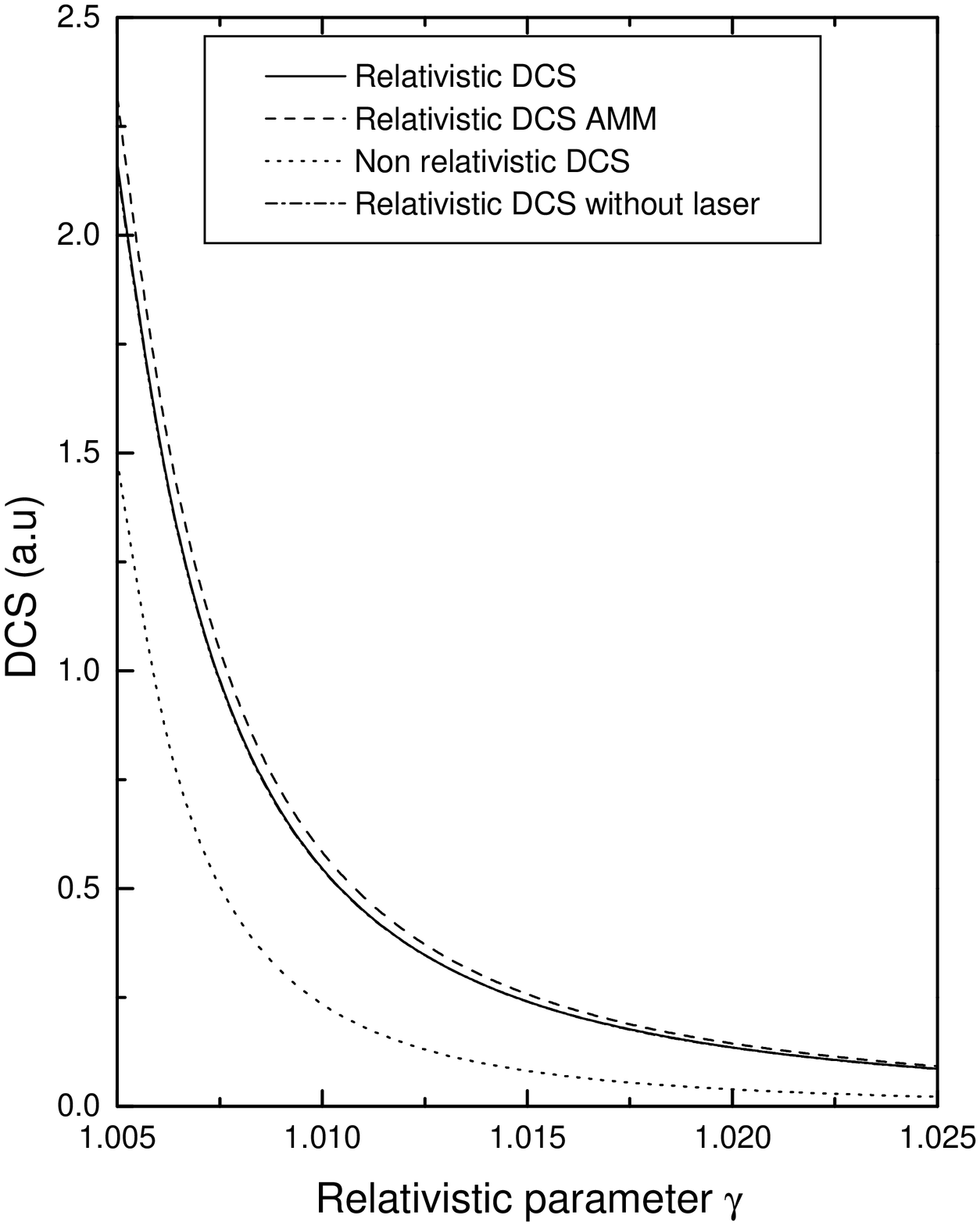}
\caption{various DCSs scaled in $10^{-5}$ as a function of relativistic parameter
$\gamma$ for an electrical field strength of $\varepsilon=0.1$
$a.u$ and an angle $\theta_{f}=135^{\circ}$. The corresponding
number of photons exchanged is $\pm1000$.}
\end{figure}
Two interesting comparisons can provide further physical insights
concerning this process. The first one concerns the ratios
$R_{1}=\ln(\left(\frac{d\sigma}{d\Omega_f}\right)^{RL}_{AMM}/\left(\frac{d\sigma}{d\Omega_f}\right)^{RNL})$
and
$R_{2}=\ln(\left(\frac{d\sigma}{d\Omega_f}\right)^{RL}/\left(\frac{d\sigma}{d\Omega_f}\right)^{RNL})$.
Maintaining the same geometry and the same values of $\varepsilon$
and $\gamma$ (that is $\varepsilon=0.05$ $a.u$ and
$\gamma=1.0053$), the upper curve for $R_{1}$ is showing a
minimum at the peak $\theta_{f}\approx33^{\circ}$ and is neatly
distinguishable from the second curve given by $R_{2}$. The fact
that these two curves have their minima located at
$\theta_{f}\approx33^{\circ}$ is not surprising since they have
been divided by $\left(d\sigma/d\Omega_f\right)^{RNL}$,
that is the relativistic DCS without a laser field. Since we have
taken the logarithm of the ratio of both DCS with an without AMM,
the violation of the pseudo sum-rule is clearly visible in Fig. 7,namely for $R_{2}$.
The behaviour of both ratios
are nearly the same from negative values of $\theta_{f}$, meaning
that for those angles, the contribution of the term
$1/2\sum_{s_{i}}\sum_{s_{f}}\left|M_{fi}^{(n)}\right|^{2}$
has an overall effect of increasing $R_{1}$ compared to that of
$R_{2}$ but the signature of the electron's AMM is not important.
Such a behaviour must be checked for the relativistic regime.
Fig.7 gives the ratios $R_{1}$ and $R_{2}$.\\
\begin{figure}[h]
\includegraphics[angle=0,width=2.5 in,height=3 in]{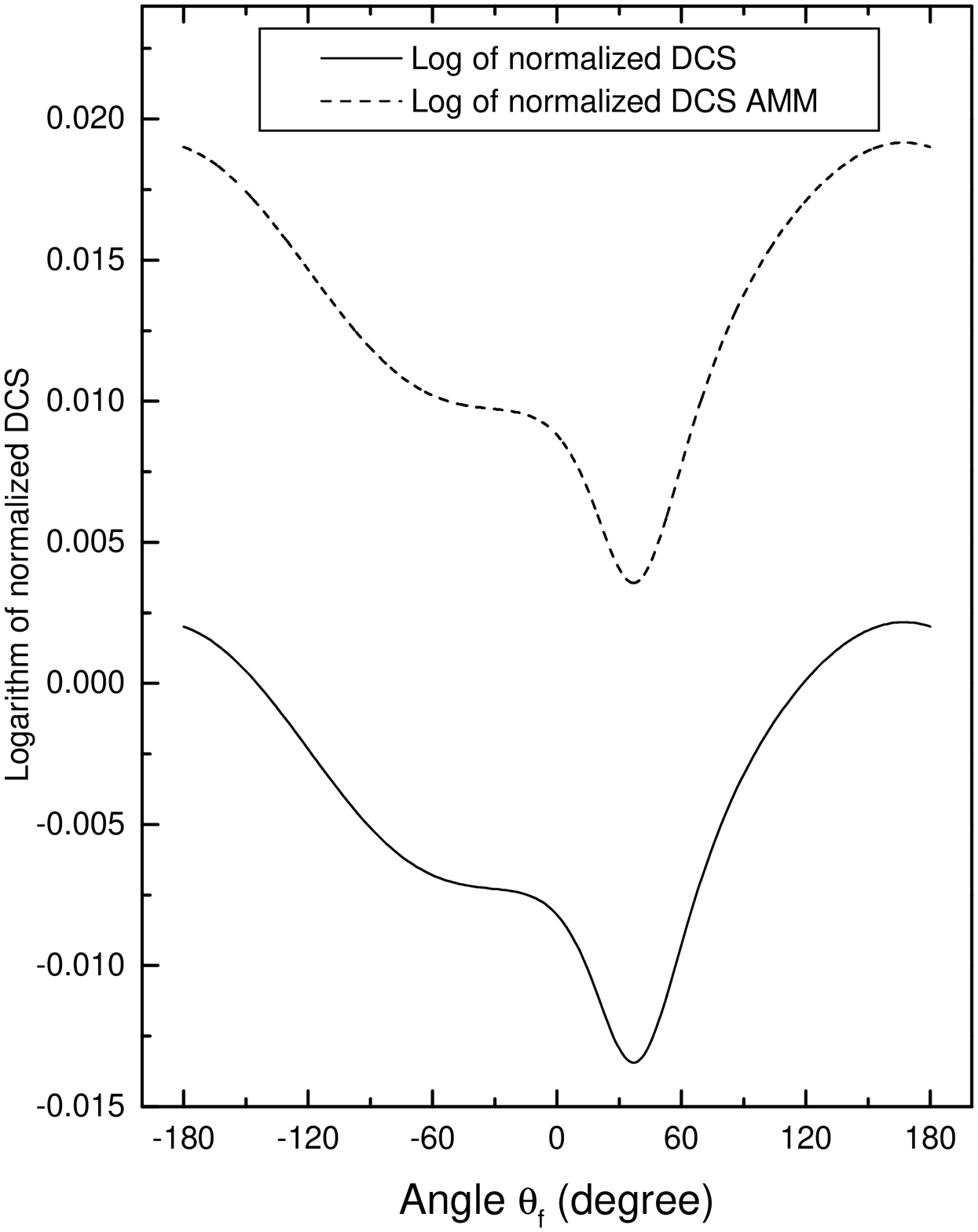}
\caption{ Logarithm of the relativistic DCS with AMM and
relativistic DCS normalized to relativistic DCS without a laser
field for an electrical field strength $\varepsilon=0.05$ $a.u$
and relativistic parameter $\gamma=1.0053$. The corresponding
number of photons exchanged is $\pm1200$.}
\end{figure}
\begin{figure}[h]
\includegraphics[angle=0,width=2.5 in,height=3 in]{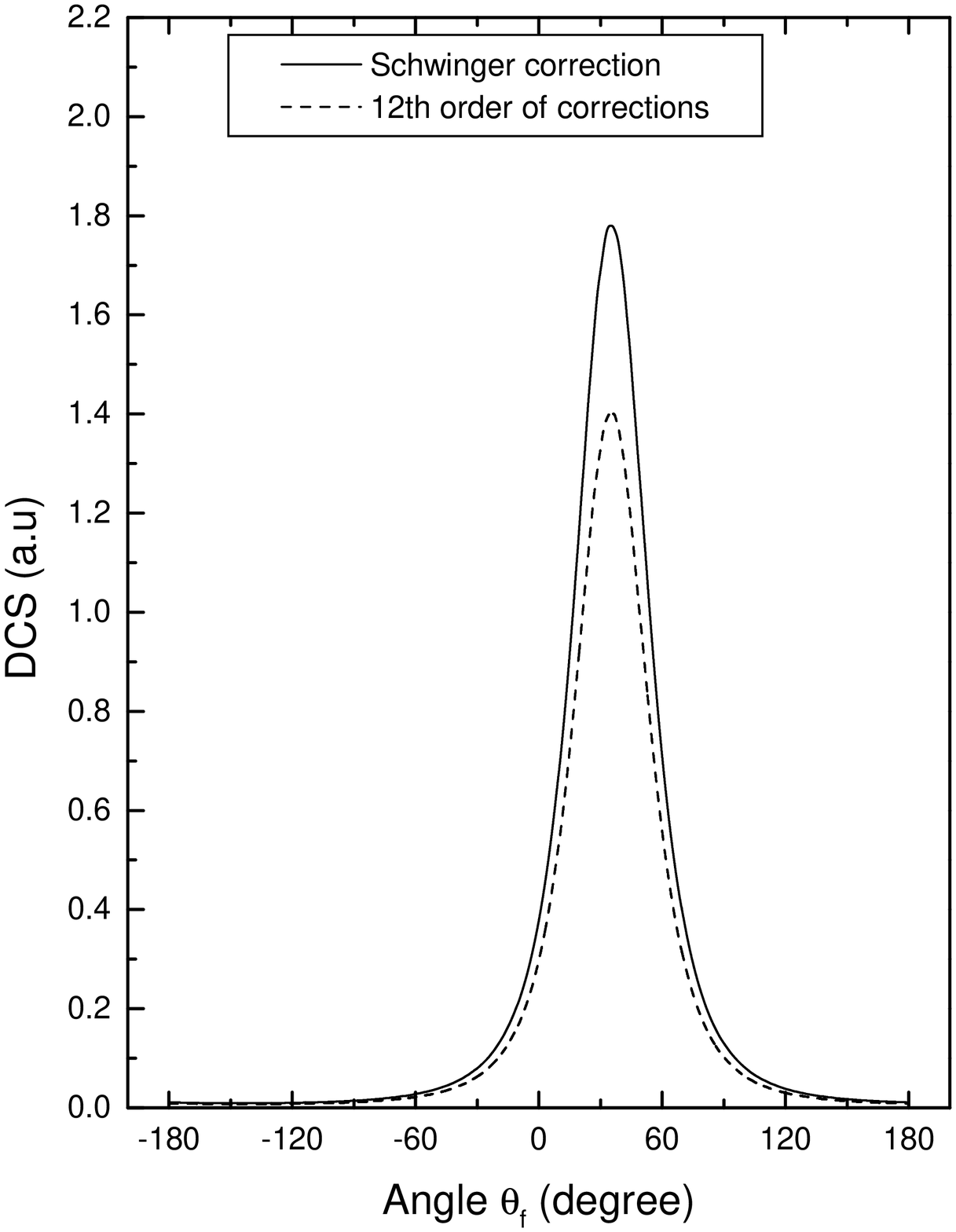}
\caption{ Comparison between DCS with AMM for Schwinger
correction and 12th order of corrections, for an electrical field
strength $\varepsilon=0.05$ $a.u$ and relativistic parameter
$\gamma=1.0053$. The DCSs are scaled in $10^{-3}$ and the corresponding number of photons exchanged is
$\pm1000$.}
\end{figure}
To end this section concerning the non relativistic regime, one
may ask whether the value used throughout this work for the
electron's anomaly is really sensitive to the order of the
radiative corrections. This is indeed the case since Fig.8 shows
that when using only the second order radiative correction found
by Schwinger \cite{2}, we have an over estimation for the
$\left(d\sigma/d\Omega_f\right)^{RL}_{AMM}$(Schwinger) of
about $28$ percent compared to the
$\left(d\sigma/d\Omega_f\right)^{RL}_{AMM}$(12th order)
meaning that radiative corrections reduce the values of the
angular distributions. We have checked this for the same geometry
and for $n=\pm1000$ photons. Needless to say that such a
comparison is out of our reach in the relativistic domain since
the number $n$ must be very high to come with a convincing
conclusion .
\subsection{The relativistic regime.}
The main difficulty when investigating the relativistic domain is
the limitation due to the computing power of our computer (namely
an intel (R) Core (TM) 2 DUO CPU  2.2 GHZ). Indeed, with such a
material, it is not possible to go beyond a certain limit for
$n$, the number of photons exchanged.\\
We have tried whenever this was possible, to extract
qualitative results that will not change drastically when $n$ is
increased.\\
\begin{figure}[h]
\includegraphics[angle=0,width=2.5 in,height=3 in]{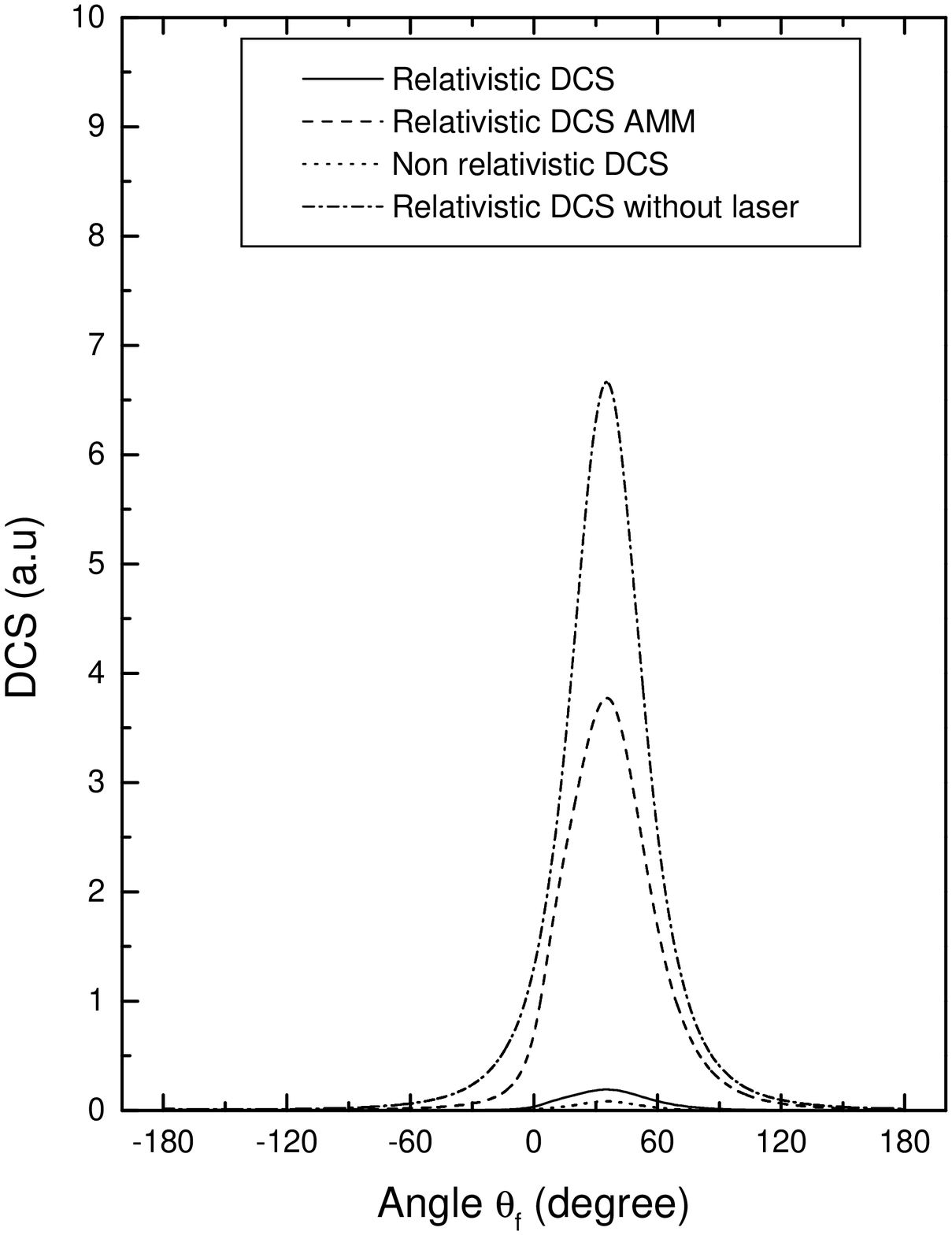}
\caption{ The various relativistic DCSs scaled in $10^{-8}$ as a function of the angle
$\theta_{f}$ in degrees for an electrical field strength of
$\varepsilon=1$ $a.u$ and a relativistic parameter $\gamma=2$. The
corresponding number of photons exchanged is $\pm5000$..}
\end{figure}
The angular parameters are the same as well as the laser
frequency whereas $\varepsilon=1$ $a.u$ and $\gamma=2$.\\
The first physical quantities to be investigated are of course the
various differential cross sections. Bearing in mind the
limitation of our computer, we can't go beyond a certain number
of photons exchanged. The first observation that can be made
concern the magnitudes of these DCSs that are strongly decreased
in the relativistic regime. The dressed momentum coordinates
($\mathbf{q}_{i}$,$\mathbf{q}_{f}$) are now noticeably different
from the non dressed momentum coordinates
($\mathbf{p}_{i}$,$\mathbf{p}_{f}$), therefore the maximum or peak of
the various DCSs is now located at nearly $29º$. In this regime,
the differences between
$\left(d\sigma/d\Omega_f\right)^{RL}_{AMM}$ and
$\left(d\sigma/d\Omega_f\right)^{RL}$ are much more
pronounced than those in the non relativistic regime and this was
to be expected because there is a strong correlation between the electric field strength and the electron's
anomaly$\kappa$. Since the former one is increased from
$\varepsilon=0.05$ $a.u$ to $\varepsilon=1$ $a.u$, the dependence
of
the DCSs are clearly shown in Fig.9.\\
This figure also shows that the overall behaviour of
$\left(d\sigma/d\Omega_f\right)^{RL}_{AMM}$ .vs.
$\left(d\sigma/d\Omega_f\right)^{RL}$ does not vary even
with increasing values of $n$, we have then turned our
investigation to an other aspect of the behaviour of a sole DCS
with respect to the number of photons exchanged.\\
Fig.10 shows the increase of
$\left(d\sigma/d\Omega_f\right)^{RL}$ for the different values of
the summation over $n$, the first summation is over $n=\pm1000$
photons while the last one is over $n=\pm5000$ photons. Since the
relativistic DCS without laser field is nearly $6.75 10^{-8}$ at
its maximum, it is obvious that one has to sum over a very large
number of photons exchanged in order to obtain at least the
$10^{-8}$ order of magnitude. Even for $n=\pm5000$, the value of
$\left(d\sigma/d\Omega_f\right)^{RL}$ is nearly $2.10^{-9}$ at
its maximum. As there is linear relation between the summation
over $n$ and the value of $\left(d\sigma/d\Omega_f\right)^{RL}$
at its peak, it is however possible to investigate this
dependence by reducing the
angular distribution interval.\\
\begin{figure}[h]
\includegraphics[angle=0,width=2.5 in,height=3 in]{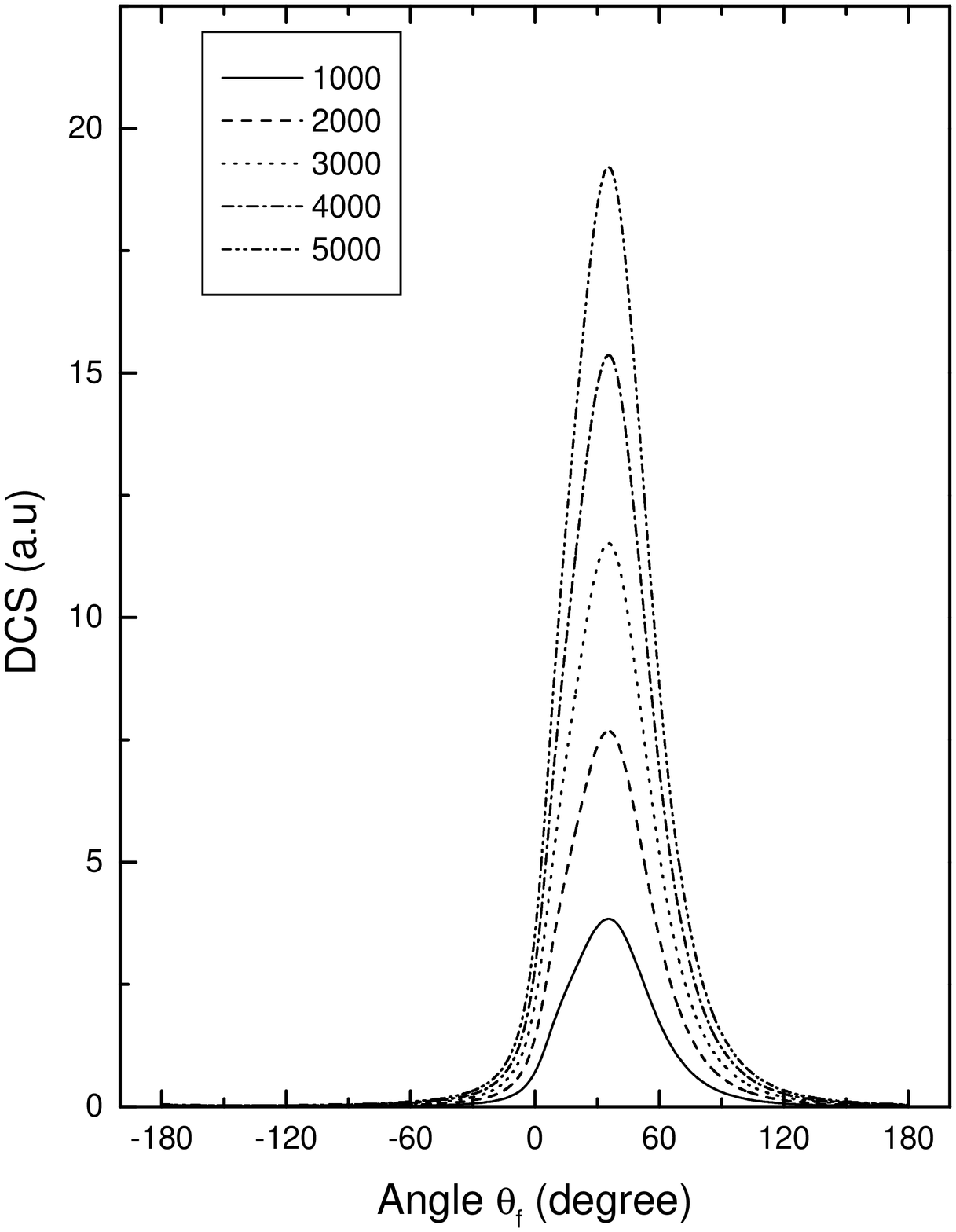}
\caption{ The relativistic DCS with a laser field  scaled in $10^{-10}$ for various
values of number of photons exchanged, as a function of the angle
$\theta_{f}$ in degrees for an electrical field strength of
$\varepsilon=1$ $a.u$ and a relativistic parameter $\gamma=2$.}
\end{figure}
In Fig.11,
\begin{figure}[h]
\includegraphics[angle=0,width=2.5 in,height=3 in]{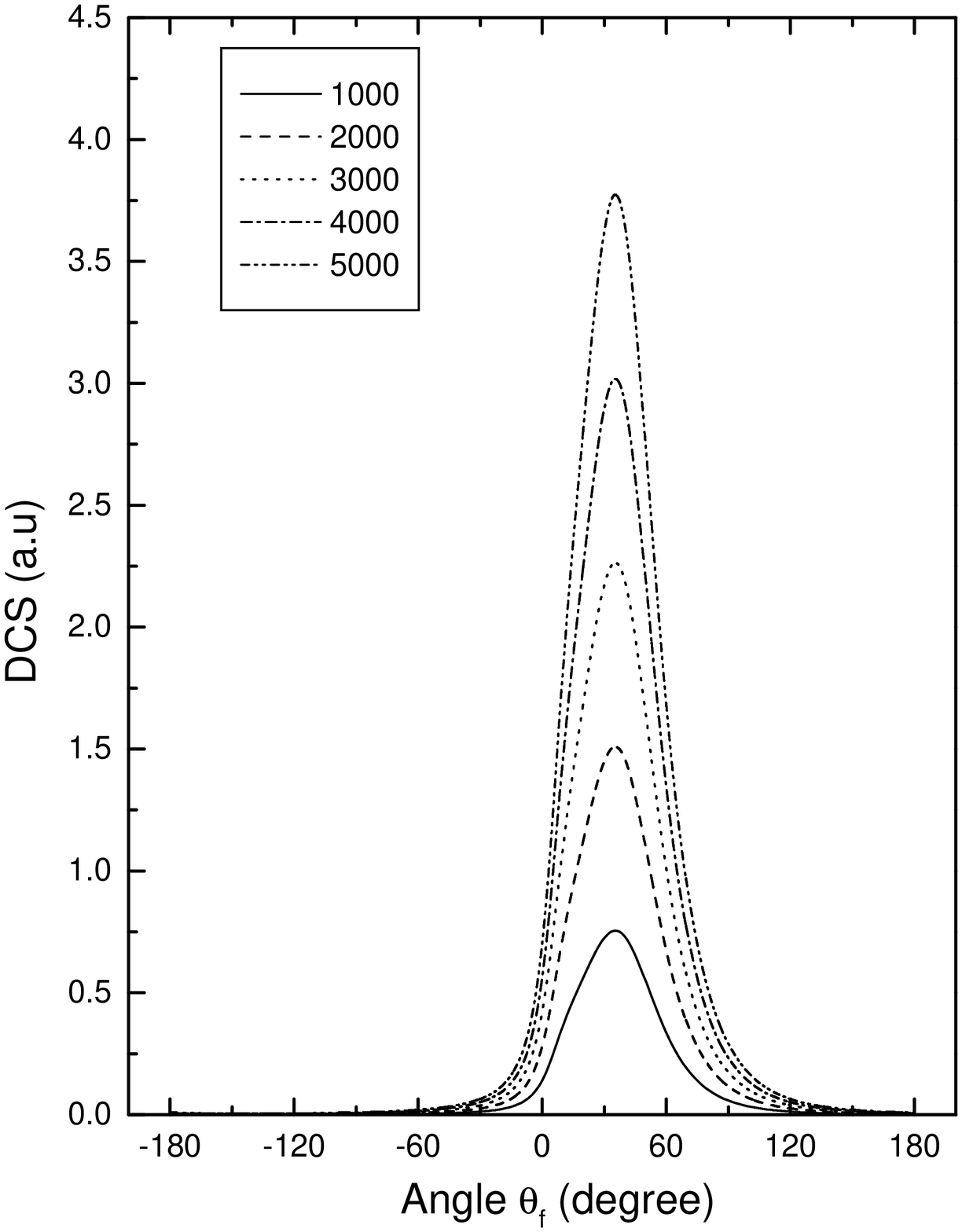}
\caption{ The relativistic DCS with AMM scaled in $10^{-8}$ for various values of
number of photons exchanged, as a function of the angle
$\theta_{f}$ in degrees for an electrical field strength of
$\varepsilon=1$ $a.u$ and a relativistic parameter $\gamma=2$.}
\end{figure}
 we show the same dependence of
$\left(d\sigma/d\Omega_f\right)^{RL}_{AMM}$ and what
emerges is a very different picture. Indeed, the value of this DCS
at its maximum is now $4.10^{-8}$ for $n=\pm5000$ which is close
to the value of $\left(d\sigma/d\Omega_f\right)^{RL}$ at
the same maximum. This means that we can, by reducing the angular
distribution interval, investigate if there is violation of the
pseudo sum rule as in the non relativistic regime. An interesting
behaviour emerges when we vary the electrical field strength from
$\varepsilon=0.05$ $a.u$ to $\varepsilon=1$ $a.u$. As $\gamma=2$,
there is an interval of values of (0.05, 0.19) where there is a
very weak quasi-linear decrease of
$\left(d\sigma/d\Omega_f\right)^{RL}$ and this DCS decrease
in a more pronounced manner in the interval (0.2, 0.3) while
smoothly decreasing of other values of $\varepsilon$. This is
consistent with the results shown in Fig.10-11 where the various
DCSs have decreased values for $\gamma=2$ and $\varepsilon=1$
$a.u$. As for the quasilinear decrease of
$\left(d\sigma/d\Omega_f\right)^{RL}$, the explanation is
rather obvious since these values of $\varepsilon$ are in the
vicinity of a "quasi non relativistic" regime as far as we
consider only $\varepsilon$. Indeed, considering the summation
over $n$ ($\pm20000$), this plateau-like behaviour means that we
have reached and overtaken the required summation over $n$ for
which the $\left(d\sigma/d\Omega_f\right)^{RL}$ converge to
$\left(d\sigma/d\Omega_f\right)^{RNL}$.\\
\begin{figure}[h]
\includegraphics[angle=0,width=2.5 in,height=3 in]{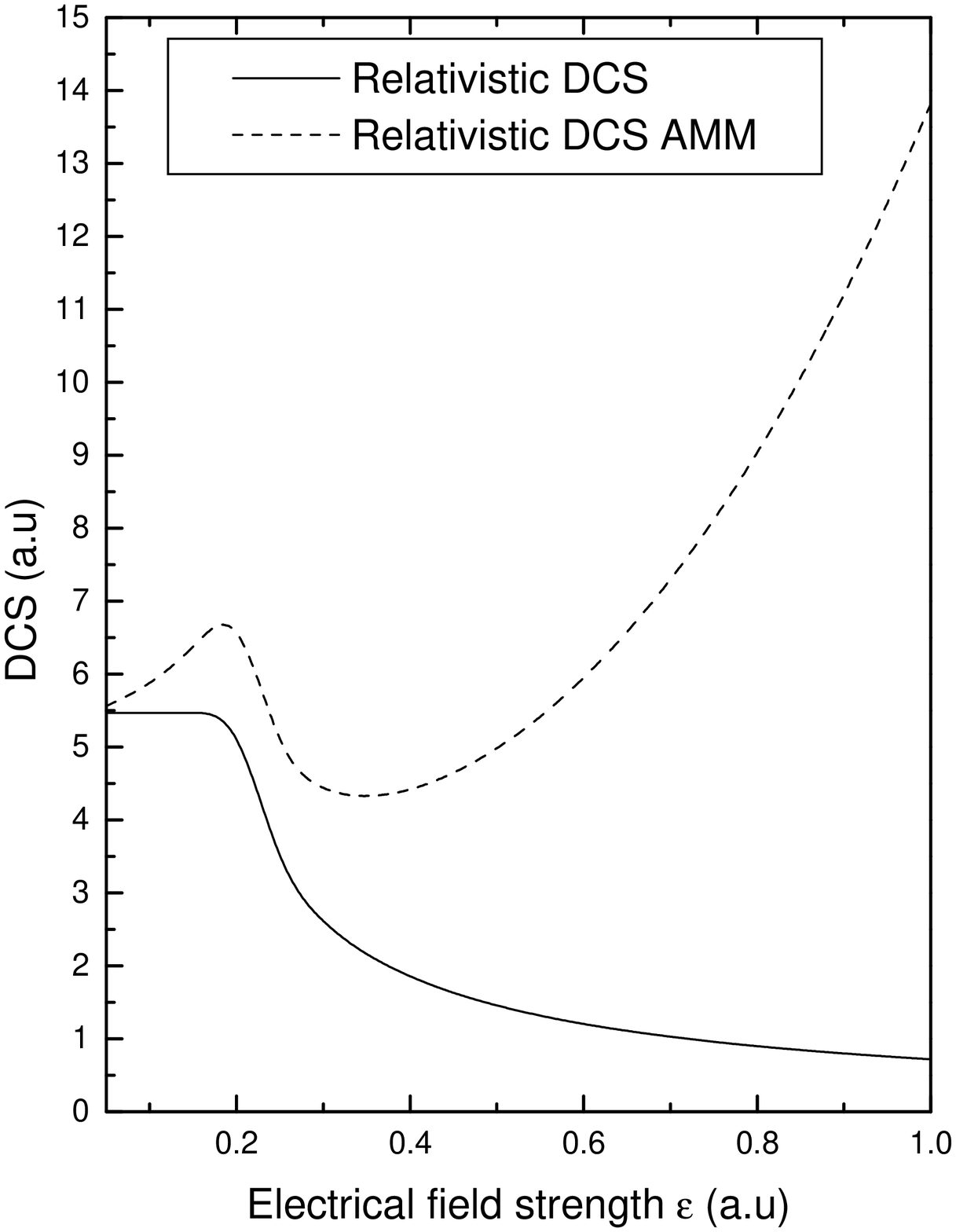}
\caption{ The two interesting DCSs  scaled in $10^{-10}$ as a
function of electrical field strength of $\varepsilon$ for a
relativistic parameter $\gamma=2$ and an angle
$\theta_{f}=135^{\circ}$. The corresponding number of photons
exchanged is $\pm20000$.}
\end{figure}
A contrasting behaviour is observed for
$\left(d\sigma/d\Omega_f\right)^{RL}_{AMM}$ where we have
summed as before over $\pm20000$ photons. Even for $\gamma=2$,
both DCSs remain close for small values of $\varepsilon$ typically
($0.05$,$0.075$) but begin to deviate from each other as
$\varepsilon$ increase. A peak of
$\left(d\sigma/d\Omega_f\right)^{RL}_{AMM}$ is present for
$\varepsilon\simeq0.19$ $a.u$ then there is a minimum for
$\varepsilon\simeq0.38$ $a.u$ and then a rapid increase for
greater values of $\varepsilon$. The latter can be easily
explained since the spinor part of
$\left(d\sigma/d\Omega_f\right)^{RL}_{AMM}$ is strongly
dependent on $\varepsilon$ as well as $\kappa$.\\
The first maximum and the next minimum are difficult to interpret
since the expression of
$\frac{1}{2}\sum_{s_{i}}\sum_{s_{f}}\left|M_{fi}^{(n)}\right|^{2}$
is very long and not prone to analytical investigation. However,
since the angular parameters are fixed, $\gamma=2$, these can
only be tracked back to the overall dependence of the spinor part
of $\left(d\sigma/d\Omega_f\right)^{RL}_{AMM}$ on the
electric field strength. These dependence with respect to
$\varepsilon$ for both DCSs are shown in Fig.12.\\
\begin{figure}[h]
\includegraphics[angle=0,width=2.5 in,height=3 in]{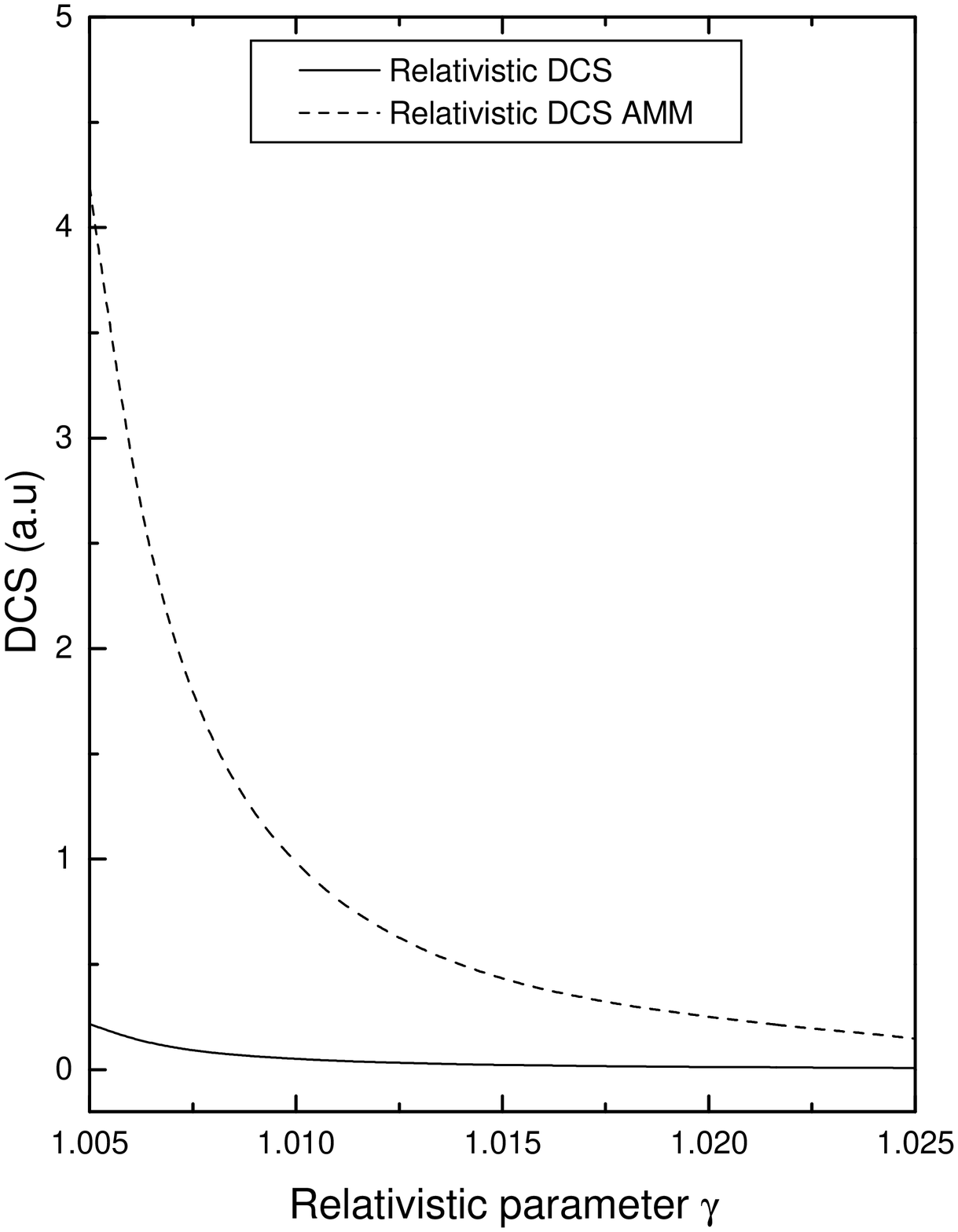}
\caption{ various DCSs  scaled in $10^{-4}$ as a function of relativistic parameter
$\gamma$ for an electrical field strength of $\varepsilon=1$ $a.u$
and an angle $\theta_{f}=135^{\circ}$. The corresponding number
of photons exchanged is $\pm10000$.}
\end{figure}
The relativistic effects on these DCSs can be investigated by
varying the relativistic parameter $\gamma$. The number of
photons exchanged is $\pm10000$. But as we have $\varepsilon=1$
$a.u$, the interpretation of the curve obtained when $\gamma$
start from $1.005$ to the value $1.025$ must be cautiously
carried out since there is an interplay between relativistic
effect due to the variation of $\gamma$ and the fixed value of
$\varepsilon$ on the one hand while on the other hand it does not
give the whole picture of what really happens when $\gamma$
varies from $1.005$ to $2$. We have remedied to this
computational task by studying  three regions for the variation
of $\gamma$, namely ($1.005$,$1.025$), ($1.495$,$1.505$),
($1.995$,$2$). An additional difficulty arises where varying
$\gamma$ (while $\varepsilon$ remains fixed, $\varepsilon=1$
$a.u$), because both DCSs decrease by a factor of magnitude $4$
(from $\sim 10^{-4}$ to $10^{-8}$) when $\gamma$ approaches $2$.
Therefore, even if $\left(d\sigma/d\Omega_f\right)^{RL}_{AMM}$ is
always greater than  $\left(d\sigma/d\Omega_f\right)^{RL}$, this
cannot be visually translated in Fig.13. However, simulations for
$\gamma\in(1.495, 1.505)$ and $\gamma\in(1.995, 2)$ give results
that are consistent with previous ones. The effect of the
frequency $\omega$ is shown in Fig.14 by varying $\omega$ from
$0.04$ $a.u$ to $0.1$ $a.u$, for a relativistic parameter
$\gamma=2$, an electric field strength $\varepsilon=1$ and for
$n=\pm10000$ photons. In this regime, the DCSs are not similar in
shape as in the non relativistic regime ( see Fig.4). While both
DCSs decrease in value (by an order of magnitude $3$),
$\left(d\sigma/d\Omega_f\right)^{RL}$ is roughly quasi-linear
whereas $\left(d\sigma/d\Omega_f\right)^{RL}_{AMM}$ decreases from
its maximum value $7.5 10^{-10}$ ($a.u$) to a minimum located at
$5.10^{-10}$ ($a.u$) and then increases.
\begin{figure}[h]
\includegraphics[angle=0,width=2.5 in,height=3 in]{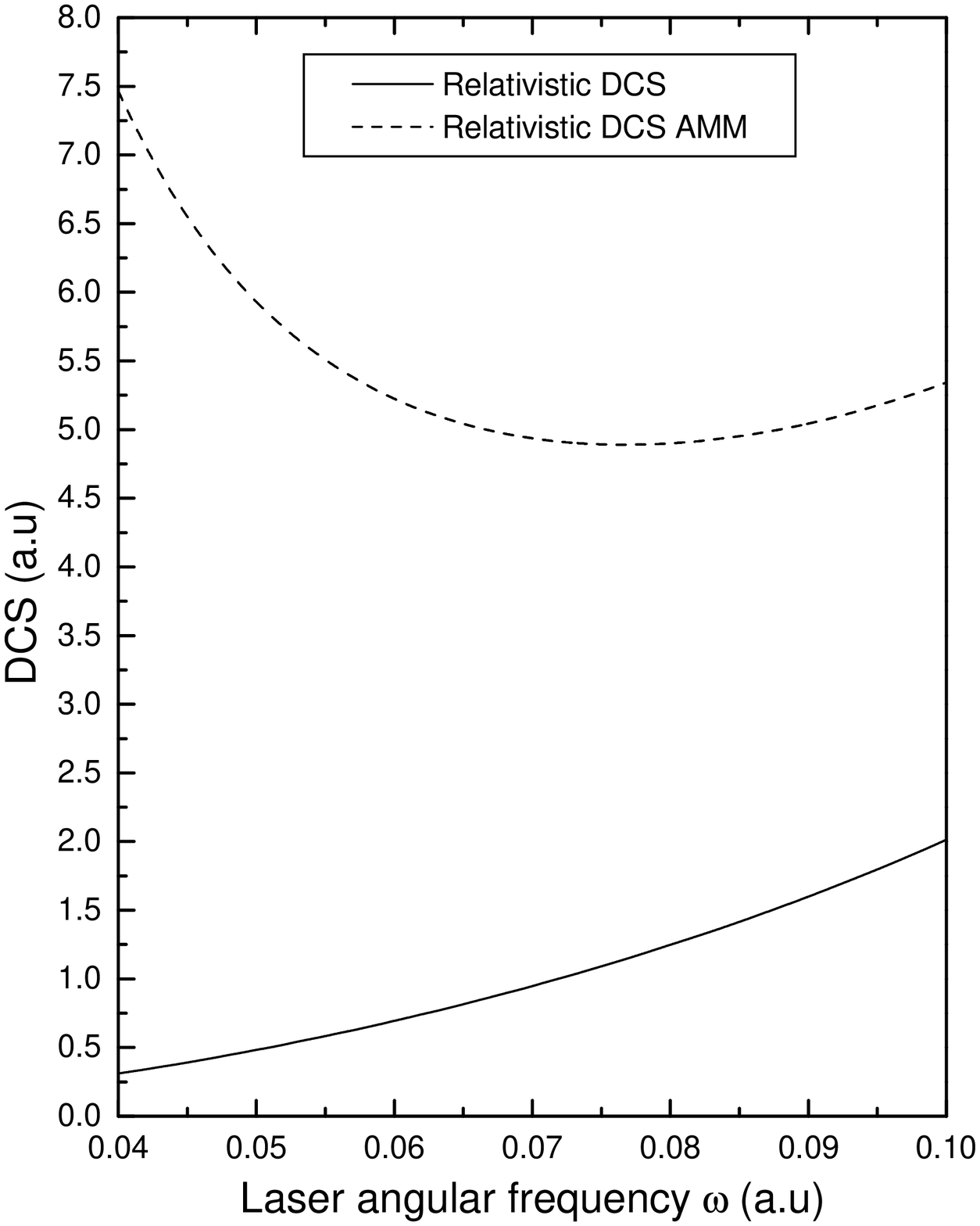}
\caption{ The various relativistic DCSs scaled in $10^{-10}$ as a
function of $\omega$ for an electrical field strength of
$\varepsilon=1$ $a.u$, a relativistic parameter $\gamma=2$ and an
angle $\theta_{f}=135^{\circ}$. The corresponding number of
photons exchanged is $\pm10000$.}
\end{figure}
\section{Conclusion}
In this work, we have presented new results concerning the effects of the electron's anomalous magnetic
moment on the process of laser-assisted electron-atomic hydrogen elastic collisions. We have used throughout this work the recent experimental value of the anomaly $a$ found by Gabrielse \cite{3}. We have focused our study on the electronic dressing with the addition of the electron anomaly. Using the Dirac-Volkov wave function that incorporates this anomaly \cite{18}, we found the analytical expression of the corresponding DCS. A spatial integral part that has been found in a previous work \cite{21} remains the same for the study of this process. The various coefficients that intervene in the expression of $S_{fi}$ have been obtained using Reduce \cite{22}. We have the same formal analogy between the DCS without and with anomaly. However, the spinor part incorporating this latter is strongly dependent on the electron's anomaly and the electric field strength.
For the non relativistic regime, the addition of the electron's AMM is noticeable but small.
When increasing the electric field strength to moderate values, this effect becomes more pronounced. For the first time, we have obtained the violation of the pseudo sum-rule \cite{23}. We have also checked that the second order correction due to J. Schwinger  \cite{2} overestimates the DCS.
In the relativistic regime, the dynamical behaviour of the DCS shows that the correlation between the terms stemming from the electron's anomaly and the electric filed strength is more pronounced even if there is an overall decrease of DCS without electron's anomaly and the DCS with the electron's anomaly.

\end{document}